\documentclass[reprint,longbibliography,amsmath,amssymb,aps,pra,nofootinbib,eqsecnum,superscriptaddress]{revtex4-1}

\usepackage{natbib}
\usepackage{array}
\usepackage{graphicx}
\usepackage{physics}
\usepackage{amsmath}
\usepackage{hyperref}
\usepackage{nicematrix,tikz}
\usepackage{comment}

\graphicspath{{./images/}}

\newcommand{\bee}{\begin{equation}}
\newcommand{\eee}{\end{equation}}
\begin{document}
\title{A quantum annealing approach to the minimum distance problem of quantum codes}

\author{Refaat Ismail}
\affiliation{Department of Physics and Astronomy, University of Kentucky, Lexington, KY, 40503 USA}

\author{Ashish Kakkar}

\author{Anatoly Dymarsky}
\email{a.dymarsky@uky.edu}
\affiliation{Department of Physics and Astronomy, University of Kentucky, Lexington, KY, 40503 USA}

\vspace{0.2cm}
\begin{abstract}
Quantum error-correcting codes (QECCs) are at the heart of fault-tolerant quantum computing. As the size of quantum platforms is expected to grow, one of the open questions is to design new optimal codes of ever-increasing size. A related challenge is to ``certify'' the quality of a given code by evaluating its minimum distance, a quantity characterizing code's capacity to preserve quantum information. This problem is known to be NP-hard. Here we propose to harness the power of contemporary quantum platforms to address this question, and in this way  help design quantum platforms of the future. 
Namely, we introduce an approach to compute the minimum distance of quantum stabilizer codes by reformulating the problem as a Quadratic Unconstrained Binary Optimization (QUBO) problem and leveraging established QUBO algorithms and heuristics as well as quantum annealing (QA) to address the latter. The reformulation as a QUBO introduces only a logarithmic multiplicative overhead in the required number of variables. We demonstrate the formulation using the D-Wave Advantage 4.1 quantum
annealer and the hybrid quantum-classical algorithm Qbsolv, and benchmark the
resulting QUBO-based implementations against specialized classical
distance-finding methods. The QA- and SA-assisted implementations show
broadly comparable performance, while the classical comparison clarifies the
relative strengths of heuristic optimization and exact distance
certification. Our results establish a solver-independent QUBO formulation
with only a logarithmic multiplicative overhead in the number of variables
and characterize connectivity and minor embedding as central limitations of
its present implementation on quantum-annealing hardware.
\end{abstract}

\maketitle

\section{Introduction}
The field of quantum error correction has been the focus of intense research in recent decades due to its role in the development of fault-tolerant quantum computing. Quantum fault tolerance offers a trade-off between efficiency in storing and manipulating information and robust protection against errors \cite{Knill_2000}. One of the central theoretical goals is to design QECCs that minimize the required resources while enhancing the protective properties.

Quantum algorithms operate on an abstract Hilbert space, known as the logical Hilbert space. When employing a QECC, we restrict the logical Hilbert space to a specific subspace of the full computational (also referred to as physical) Hilbert space. Errors that send the computation outside the code subspace could be detected and corrected. In contrast, errors that map the code subspace into itself are undetectable. The code's minimum distance $d$ is defined as the smallest weight of an undetectable error, i.e.~how many physical qubits this  error affects \cite{Gottesman_1997}.

A QECC can be characterized by three parameters: the number of physical qubits $n$, the number of logical states $K$ it protects and its minimum distance $d$. Corresponding codes are conventionally denoted as $((n, K, d))$. Constructing QECCs of ever-increasing $n$, with a fixed ratio $\ln K/n$ and the best (largest) $d/n$ is an open problem. Our emphasis will be on stabilizer codes, the most common class of quantum codes introduced in \cite{Gottesman_1997} and widely considered the best strategy to achieve fault tolerance in the near term.
A binary stabilizer code, i.e.~over qubits, protects the $K=2^k$ states, where $k$ denotes the number of logical qubits and is denoted by $[[n,k,d]]_2$. Various other code constructions have been presented \cite{Cross_2009,Breuckmann_2021}, including those generated by computational methods \cite{Cao_2022,Li_2017, yu2007graphicalquantumerrorcorrectingcodes}.

An optimal  code is the one with the highest known value of $d$ for given $n$ and $k$. In practice to certify that code is optimal or close to optimal is a difficult problem.  
Depending on how the code is defined, determining $d$ could be a computationally challenging task. In fact, finding $d$ for a general (binary) stabilizer code specified by its generating matrix is known to be NP-hard. Even approximating the minimum distance, with a guaranteed precision specified by a given multiplicative or additive factor, is also NP-hard \cite{Diagonaldistanceofquantumcodes}.  There exist multiple approaches for computing the minimum distance of quantum codes.  One of the best deterministic algorithms was developed by G.~White and M.~Grassl and has been incorporated into MAGMA \cite{White_2006, BOSMA1997235}. Multiple additional techniques have been also tailored for specific families of quantum codes, such as ~Low-Density-Parity-Check (qLDPC) codes \cite{Pryadko_2022,Dumer_2014,Dumer_verification}. 

 In this paper, we introduce a new computational technique to evaluate or bound from above the minimum distance of binary quantum stabilizer codes, by reformulating the question as a Quadratic Unconstrained Binary Optimization (QUBO) problem. A QUBO is a general framework to represent combinatorial optimization problem as a problem to minimize quadratic functions of binary variables \cite{Glover_2018}. To reformulate the minimum distance problem as a QUBO, one needs to introduce auxiliary variables, e.g.~to implement mod $2$ arithmetic. As a result, our method involves a multiplicative logarithmic overhead in terms of the total number of variables.
 
Reformulating the problem in terms of QUBO introduces a promising pathway to harness power of quantum computational schemes such  as Adiabatic Quantum Computing (AQC) and Quantum Annealing (QA), alongside a spectrum of established classical algorithms and heuristics designed specifically for QUBO problems. There is an ongoing effort to reformulate many NP-hard problems of theoretical and practical value within a unified framework of QUBO, with such problems as SAT, Knapsack, Maximum Cut, and Graph Coloring already being worked out  in the literature \cite{Glover_2018,jun2022qubo, carvalho2022qubo}. These problems arise in different fields, from Quantum Information to Machine Learning to RNA folding \cite{Date_2021, dury2020qubo, mellaerts2023quantuminspired,Zaborniak_2022}. Our work can be seen as contributing to  this  effort.  

To investigate the efficiency of our approach, implemented on contemporary quantum hardware, we focus on quantum annealing (QA) as a method to solve QUBO. QA is a quantum heuristic approach, inspired by the principles of Adiabatic Quantum Computing (AQC) \cite{AdiabaticQuantumComputingandQuantumAnnealing, Farhi_2001}. QA provides a rigorous upper bound for optimization problems, albeit without a guarantee on how close the obtained result is to the optimal value. 
To investigate the viability of QA in our case, we tested our method using both the D-Wave Advantage 4.1 quantum annealer and Qbsolv, a hybrid quantum-classical algorithm developed by D-Wave and running on their platform \cite{Okada_2019}. As a benchmark, we compared the performance of quantum and quantum-classical methods with simulated annealing (SA), a well-established classical heuristic approach \cite{doi:10.1126/science.220.4598.671}. Whenever available, we compared the results of quantum, classical, or hybrid optimization with the exact results. 

Our results show broadly comparable performance between the QA- and
SA-assisted QUBO implementations, while specialized classical
distance-finding methods remain more effective on the instances considered
here. The quantum demonstrations therefore serve primarily as a
proof-of-concept implementation of the QUBO formulation and as a study of
its present hardware limitations.

The paper is organized as follows. In Section II, we introduce the basics of stabilizer codes. We formulate the main algorithm to evaluate $d$ as a QUBO problem in Section III. Section IV provides a brief overview of AQC and QA. Section V details the comparison of purely classical, quantum, and hybrid approaches. An open-source implementation of our algorithm is available on GitHub \cite{github_code}. We conclude with a discussion in section VI.

\section{Theoretical Preliminaries}

\subsection{Stabilizer Codes: a brief overview}

One of the most frequently used  class of QECCs is the stabilizer codes. If S is an Abelian subgroup of the Pauli group $\mathcal{P}_n$\footnote{The $n$ qubit Pauli group, $\mathcal{P}_n$,  consists of all n-fold tensor products of  the Pauli operators $I, X, Y, Z$, with an overall phase $i^k$ with an integer $k$.} with $(n-k)$ generators and $-I \notin S$, then the simultaneous +1 eigenspace, $V_S$, of all elements of $S$ is an $[[n,k,d]]$ quantum stabilizer code, where $n$, $k$, and $d$ denote the number of physical qubits, logical qubits, and the code minimum distance, correspondingly. Note that the two conditions on the subgroup, being Abelian and $-I \notin S$, are to ensure that the code space is non-trivial.

To specify the stabilizer code, it is sufficient to provide $(n-k)$ generators of the stabilizer group $S$. The stabilizer generators contain essential information about the error-correcting capabilities of the code, eliminating the need to construct the code space $V_S$ explicitly. Stabilizer codes are also known as additive quantum codes due to their one-to-one correspondence with trace hermitian self-orthogonal additive classical codes over GF(4) \cite{Calderbank_1996}.

Pauli group elements can be represented as binary vectors, allowing for an efficient representation of the stabilizer group. Namely, up to a phase, any Pauli group element can be expressed as a product of X and Z operators, each raised to a power of either 0 or 1,
\begin{align}
    g \propto \left( X_1^{\alpha_1} X_2^{\alpha_2}\ldots X_n^{\alpha_n} \right) \cdot \left( Z_1^{\beta_1} Z_2^{\beta_2}\ldots Z_n^{\beta_n} \right), \quad g \in \mathcal{P}_n
    \label{eq:stabilizerfromvec}
\end{align}
Here $\alpha_i$ and $\beta_i$ are binary variables.  Combining  them together, we can represent a Pauli element using a binary vector $\boldsymbol{c}=(\boldsymbol{\alpha}|\boldsymbol{\beta})$\footnote{We use bold symbols to represent vectors.}. Multiplying two Paulis then corresponds to adding their respective binary vectors mod 2.

In this representation, the stabilizer group generators can be specified by a binary $(n-k) \times 2n$ matrix, where each row is vector representation of one of the $(n-k)$ generators. This matrix H is called the stabilizer matrix (also known as the parity check matrix). We can generate  binary representations of all elements of the stabilizer group as linear combinations (mod 2) of the generators of the group,
\begin{equation}
\begin{aligned}
\boldsymbol{c} = &H^T\boldsymbol{x} \pmod 2, \\
\boldsymbol{x} \in \mathbb{Z}_2^{\,n-k}, \quad
\boldsymbol{c} &\in \mathbb{Z}_2^{\,2n}, \quad
\mathbb{Z}_2 \equiv \{0,1\}.
\end{aligned}
\end{equation}
where $\boldsymbol{c}$ is the binary representation of a given stabilizer group element, sometimes called a  ``codeword.''

The normalizer of S, denoted $N_{\mathcal{P}_n}(S)$, consists of all elements of Pauli group that map the stabilizer group to itself under conjugation, $N_{\mathcal{P}_n}(S) = \{g \in \mathcal{P}_n \ | \ gSg^{-1}= S \}$. It is clear that $S \subseteq N_{\mathcal{P}_n}(S)$.  The operators present in the normalizer but not in stabilizer represent logical Pauli operators on the $k$ encoded logical qubits. These operators map states from the code subspace into each other, which makes them undetectable errors. 

The generators of the normalizer group, $N_{\mathcal{P}_n}(S)$, can be represented as columns of the normalizer matrix G (also known as the generator matrix), which is a $(2n) \times (n+k)$ matrix of full rank that satisfies 
\begin{equation}
\label{HG}
    H\, \Lambda \, G = 0 \ (\text{mod 2}), \quad \Lambda := 
 \begin{pmatrix}
    0 & I_n\\I_n & 0
\end{pmatrix},
\end{equation} 
with $I_n$ being the $n \times n$ identity matrix. The full rank of G ensures that all generators of the normalizer are linearly independent, while the condition \eqref{HG} ensures that all normalizer elements commute with the stabilizer group. Matrix $G$ also satisfies   ${\rm dim}\ker\left(G^T \Lambda G \right) = n-k$.

\subsection{Minimum distance} 

The weight of a Pauli operator -- an element from the Pauli group -- is the number of qubits on which it acts non-trivially. The weight of a Pauli operator can be calculated using the binary representation $\boldsymbol{c} = (\boldsymbol{\alpha} | \boldsymbol{\beta}) $ as follows,
\begin{equation}
\begin{split}
w(\boldsymbol{c}) = \boldsymbol{\alpha} \cdot \boldsymbol{\alpha} + \boldsymbol{\beta} \cdot \boldsymbol{\beta} - \boldsymbol{\alpha} \cdot \boldsymbol{\beta}.
\end{split}
\label{pauli_weight}
\end{equation}
The minimum distance of a $(k \neq 0)$ stabilizer code is the smallest weight of an operator in the set of all logical Pauli operators, i.e., the set of elements in the normalizer $N_{\mathcal{P}_n}(S)$ but not in the stabilizer $S$. 

The code minimum distance $d$ can then be expressed in terms of the normalizer matrix, $G$,  
\begin{equation}
\begin{aligned}
d = &\min_{\boldsymbol{c}} w(\boldsymbol{c}), \\
\boldsymbol{c} = G\, \boldsymbol{x} \pmod 2, \quad 
\boldsymbol{x} &\in \mathbb{Z}_2^{n+k}, \quad
\boldsymbol{c} \notin \ker(G^T\Lambda).
\end{aligned}
\label{distance_non_dual}
\end{equation}
Here the constraint $\boldsymbol{c} \notin \text{ker}(G^T\Lambda)$ is equivalent to $\boldsymbol{c} \notin \text{Im}(H^T)$ which ensures that we minimize only over the set of logical Pauli operations. A stabilizer code with minimum distance $d$ can correct all errors of weight $\lfloor \frac{d-1}{2} \rfloor$ or less, and can also correct all erasure errors\footnote{Erasure errors are errors in which the qubits affected by the error are known.} of weight $d-1$ or less \cite{Gottesman_1997}.

\subsection{Self-dual stabilizer codes}

Self-dual codes represent a special class of stabilizer codes with $k=0$. In this case the code space $V_S$  is one-dimensional, often referred to as a stabilizer state. These codes cannot store or be used to manipulate quantum information, instead, they are protocols for error detection.  By initializing a quantum processor in a stabilizer state and evolving in time such that it should return to the original state,  we can detect and identify any errors that may have accumulated during this evolution. For self-dual codes, the minimum distance of the code is the smallest weight among the weights of all non-trivial elements of the stabilizer group,
\begin{equation}
\begin{aligned}
d =& \min_{\boldsymbol{c}} w(\boldsymbol{c}), \\
\boldsymbol{c} = G\boldsymbol{x} \ \text{(mod 2)}, \quad & \boldsymbol{x} \in \mathbb{Z}_2^{n}, \quad \boldsymbol{c} \neq 0^{2n}.
\label{distance_dual}
\end{aligned}
\end{equation}
For self-dual codes, we can chose $G = H^T$. An $[[n,0,d]]$ code can detect all errors affecting up to $d-1$ qubits, and can identify all errors of weight $\lfloor \frac{d-1}{2} \rfloor$ or less \cite{Gottesman_1997}.

\section{Algorithm}
\label{sec:algorithm}

In this section, we reformulate the problem of finding the minimum distance of a stabilizer code as a QUBO
(quadratic unconstrained binary optimization)  problem.

\subsection{Quadratic binary cost function}
The transpose of the normalizer matrix can be partitioned as $G^T=\begin{pmatrix} A & B \end{pmatrix}$, where $A$ and $B$ are $n \times (n+k) $ binary matrices corresponding to Pauli X and Pauli Z support, respectively. Throughout this section, matrix
multiplication is performed using ordinary integer arithmetic. We denote the parity of an integer $m$ by \begin{equation} \mathbb{P}(m):=m\bmod 2, \label{parity_operator} \end{equation} and apply $\mathbb{P}$ componentwise to vectors. The binary vectors specifying the $X$ and $Z$ support of the Pauli operator generated by $\boldsymbol{x}\in\mathbb{Z}_2^{n+k}$ are therefore \begin{equation} \boldsymbol{\alpha} = \mathbb{P}(A\boldsymbol{x}), \qquad \boldsymbol{\beta} = \mathbb{P}(B\boldsymbol{x}). \label{alpha_beta_definition} \end{equation}

Using the definition of the Pauli weight in Eq.~\eqref{pauli_weight}, the minimum distance in Eq.~\eqref{distance_non_dual} can be written as \begin{eqnarray} d &=& \min_{\boldsymbol{x}\notin\mathcal{C}} \sum_{i=1}^{n} \left( \alpha_i^2+\beta_i^2-\alpha_i\beta_i \right), \qquad \boldsymbol{x}\in\mathbb{Z}_2^{n+k}, \label{min_problem} \\ \mathcal{C} &=& \left\{ \begin{array}{ll} \{\boldsymbol{0}\}, & k=0,\\[2mm] \left\{ \boldsymbol{x}\;|\; G^T\Lambda G\boldsymbol{x}=\boldsymbol{0} \right\}, & k\neq 0. \end{array} \right. \label{optimization_prob} \end{eqnarray} The minimization in Eq.~\eqref{min_problem} already has a quadratic cost function in the binary variables $\alpha_i$ and $\beta_i$. It differs from a QUBO in two respects: these variables are parities of integer sums, and the allowed configurations of $\boldsymbol{x}$ are constrained by $\boldsymbol{x}\notin\mathcal{C}$. We address these two issues separately, beginning with the parity terms.

\subsection{Reformulation of the mod 2 summation}
\label{sec:mod2_reformulation}

The cost function \eqref{min_problem} is quadratic in the parities $\alpha_i$ and
$\beta_i$, so everything hinges on how a parity is written in terms of the binary
variables $x_j$. We can express it one of two ways, which
fail to be QUBO-compatible for opposite reasons:

\begin{equation}
\alpha_i
=
\sum_j A_{ij}x_j \bmod 2
=
\frac{1}{2}
\left[
1-\prod_j\left(1-2A_{ij}x_j\right)
\right].
\label{parity_forms}
\end{equation}
The first form is linear in the $x_j$ but contains a
modular operation, which a QUBO does not allow. The second uses the right arithmetic,
but expanding the product produces monomials of degree up to the row weight
$w_i = \sum_j A_{ij}$, whereas a QUBO admits nothing beyond quadratic. We therefore develop a different representation that removes the modular arithmetic while preserving quadraticity.

\paragraph{Decomposition of the Pauli weight into parities.}

The first step is to rewrite the Pauli-weight contribution in a form that involves only sums of parity functions. The original expression \begin{equation} \alpha_i^2+\beta_i^2-\alpha_i\beta_i \end{equation} contains the product $\alpha_i\beta_i$ of two parities. Instead, the same quantity can be expressed as a sum of three parities: the parity associated with the $X$ support, the parity associated with the $Z$ support, and the parity of their sum.

Since $\alpha_i,\beta_i\in\{0,1\}$, we have $\alpha_i^2=\alpha_i$, $\beta_i^2=\beta_i$, and 
\begin{equation} \mathbb{P}(\alpha_i+\beta_i) = \alpha_i+\beta_i-2\alpha_i\beta_i. \label{xor_identity} \end{equation}
It follows that 
\begin{equation} \alpha_i^2+\beta_i^2-\alpha_i\beta_i = \frac{1}{2} \left[ \alpha_i+\beta_i+ \mathbb{P}(\alpha_i+\beta_i) \right]. \label{weight_parity_identity} \end{equation}
Using Eq.~\eqref{alpha_beta_definition}, the third parity can be written as 
\begin{equation} \begin{aligned} \mathbb{P}(\alpha_i+\beta_i) &= \mathbb{P} \left[ \mathbb{P}\left((A\boldsymbol{x})_i\right) + \mathbb{P}\left((B\boldsymbol{x})_i\right) \right] \\ &= \mathbb{P} \left( (A\boldsymbol{x})_i + (B\boldsymbol{x})_i \right) \\ &= \mathbb{P} \left( \left[(A+B)\boldsymbol{x}\right]_i \right). \end{aligned} \label{third_parity} \end{equation}
Hence the Pauli-weight contribution at qubit $i$ becomes 
\begin{equation}
\begin{aligned}
\alpha_i^2+\beta_i^2-\alpha_i\beta_i
=
\frac{1}{2}\Big[
&
\mathbb{P}\left((A\boldsymbol{x})_i\right)
+
\mathbb{P}\left((B\boldsymbol{x})_i\right)
\\
&
+
\mathbb{P}\left(\left[(A+B)\boldsymbol{x}\right]_i\right)
\Big].
\end{aligned}
\label{three_parities}
\end{equation}
Thus, the product of the two original parities has been replaced by three parity functions of the same form, associated respectively with $A$, $B$, and $A+B$. To treat these three terms uniformly, we define 
\begin{equation} C^{(1)}:=A, \qquad C^{(2)}:=B, \qquad C^{(3)}:=A+B, \label{define_Cr} \end{equation} 
and 
\begin{equation} m_i^{(r)} := \left(C^{(r)}\boldsymbol{x}\right)_i, \qquad r=1,2,3. \label{define_mir} \end{equation}
Equation~\eqref{three_parities} then takes the compact form 
\begin{equation} \alpha_i^2+\beta_i^2-\alpha_i\beta_i = \frac{1}{2} \sum_{r=1}^{3} \mathbb{P}\left(m_i^{(r)}\right). \label{parity_identity} \end{equation}
It remains to eliminate the parity function itself.

\paragraph{Parity as a quadratic minimization.} 
For any nonnegative integer $m$, its parity can be written  as 
\begin{equation} \mathbb{P}(m) = \min_{q\in\mathbb{Z}_{\geq 0}} (m-2q)^2. \label{parity_minimization} \end{equation}
Indeed, if $m$ is even, the minimum is attained for $q=m/2$ and is equal to zero. If $m$ is odd, $m-2q$ is odd for every integer $q$, and its smallest absolute value is one. The minimum is therefore one. Applying Eq.~\eqref{parity_minimization} to each term in Eq.~\eqref{parity_identity} gives 
\begin{equation} d = \min_{\substack{ \boldsymbol{x}\notin\mathcal{C}\\ \{q_i^{(r)}\} }} \frac{1}{2} \sum_{i=1}^{n} \sum_{r=1}^{3} \left( m_i^{(r)}-2q_i^{(r)} \right)^2. \label{integer_quadratic} \end{equation}

Thus, the modular operations have been replaced by a minimization over auxiliary nonnegative integer variables $q_i^{(r)}$. Since $m_i^{(r)}$ is linear in $\boldsymbol{x}$, each term in Eq.~\eqref{integer_quadratic} is already quadratic once the auxiliary integers are represented by binary variables. 

\paragraph{Binary encoding of the auxiliary integers.} 

The range required for each $q_i^{(r)}$ follows from the largest value of
$m_i^{(r)}$. Define
\begin{equation}
M_r
:=
\max_i\sum_j C_{ij}^{(r)},
\qquad
r=1,2,3.
\label{define_Mr}
\end{equation}
Since $\boldsymbol{x}$ is binary,
\begin{equation}
0\leq m_i^{(r)}\leq M_r,
\end{equation}
and the minimizing integer in Eq.~\eqref{parity_minimization} never needs to
exceed $\lfloor M_r/2\rfloor$. We therefore encode
\begin{equation}
q_i^{(r)}
=
\sum_{\ell=0}^{s_r-1}
2^\ell q_{i\ell}^{(r)},
\qquad
q_{i\ell}^{(r)}\in\mathbb{Z}_2,
\label{binary_q}
\end{equation}
where
\begin{equation}
s_r
=
\left\lceil
\log_2\left(
\left\lfloor\frac{M_r}{2}\right\rfloor+1
\right)
\right\rceil.
\label{define_sr}
\end{equation}
Substituting Eq.~\eqref{binary_q} into Eq.~\eqref{integer_quadratic} makes
every term quadratic in binary variables. Collectively denoting the auxiliary
binary variables by $\boldsymbol{y}$, the weight objective can therefore be
written as
\begin{equation}
F(\boldsymbol{x},\boldsymbol{y})
=
\begin{pmatrix}
\boldsymbol{x}\\
\boldsymbol{y}
\end{pmatrix}^{T}
Q
\begin{pmatrix}
\boldsymbol{x}\\
\boldsymbol{y}
\end{pmatrix}.
\label{qubo_objective}
\end{equation}

The explicit block form of $Q$ is given in Appendix~\ref{app:assembly}. The number of auxiliary binary variables introduced by the parity reformulation is 
\begin{equation} N_{\mathrm{par}} = n\sum_{r=1}^{3}s_r. \label{number_parity_aux} \end{equation} Since 
\begin{equation} M_1,M_2\leq n+k, \qquad M_3\leq 2(n+k), \end{equation} each $s_r$ grows logarithmically with $n+k$, and hence 
\begin{equation} N_{\mathrm{par}} = O\left(n\log(n+k)\right). \end{equation} Because $k\leq n$, the total number of binary variables scales as \begin{equation} O\left(n\log n\right). \end{equation}

\subsubsection{Self-dual codes} 
\label{sec:self_dual} 

For self-dual codes, the number of auxiliary variables can be reduced further. Every self-dual code is equivalent to a graph code \cite{Danielsen_2006},
for which the normalizer matrix can be chosen such that
\begin{equation} G^T = \begin{pmatrix} I_n & B \end{pmatrix}, \label{graph_code_form} \end{equation}
where $B$ is the adjacency matrix of a simple undirected graph. In this representation, 
\begin{equation} \alpha_i=x_i, \qquad \beta_i=\mathbb{P}\left((B\boldsymbol{x})_i\right). \end{equation}
Because the $X$ support is now given directly by $x_i$, the three-parity
construction of the previous subsection can be replaced by a more economical
representation involving only one family of auxiliary integers. Define \begin{equation} m_i:=(B\boldsymbol{x})_i. \label{selfdual_m} \end{equation} The local Pauli-weight contribution can be written as \begin{equation} x_i^2+\beta_i^2-x_i\beta_i = x_i + \min_{q_i\in\mathbb{Z}_{\geq0}} (m_i-2q_i)(m_i-2q_i-x_i). \label{selfdual_identity} \end{equation} For $x_i=0$, the minimization reduces to Eq.~\eqref{parity_minimization} and returns $\beta_i$. For $x_i=1$, its minimum is zero independently of the parity of $m_i$, and the expression returns one. Equation~\eqref{selfdual_identity} therefore reproduces the Pauli weight in both cases. The minimum-distance problem becomes \begin{equation} d = \min_{\substack{ \boldsymbol{x}\neq\boldsymbol{0}\\ \{q_i\} }} \sum_{i=1}^{n} \left[ x_i + (m_i-2q_i)(m_i-2q_i-x_i) \right]. \label{selfdual_quadratic} \end{equation} The integers $q_i$ are encoded as in Eq.~\eqref{binary_q}. Defining \begin{equation} M_B:=\max_i\sum_j B_{ij}, \qquad s_B = \left\lceil \log_2 \left( \left\lfloor\frac{M_B}{2}\right\rfloor+1 \right) \right\rceil, \end{equation} this construction requires only \begin{equation} N_{\mathrm{par}}^{\mathrm{SD}} = ns_B \end{equation} auxiliary binary variables. The corresponding QUBO matrix is given in Appendix~\ref{app:assembly_selfdual}.

\subsection{Reformulating the constraints} 
\label{sec:constraints} 
The construction above removes the modular arithmetic from the objective, but the minimization is still restricted by $\boldsymbol{x}\notin\mathcal{C}$. To obtain an unconstrained binary optimization problem, this remaining condition must also be eliminated. We consider two ways of doing so. 

\subsubsection{Constraint as a penalty term} 
\label{sec:constraint_penalty} 

Different stabilizer codes are called equivalent if their stabilizer generators can be mapped to each other using conjugation by local Clifford operations\footnote{Local Clifford operations are tensor products of single-qubit unitaries  that map Pauli operators to themselves under conjugation.}, and qubit permutations. The definition of code equivalence ensures that equivalent codes have the same minimum distance. Code equivalences, together with the freedom to select generators of the normalizer, can be used to bring the normalizer  matrix $G$ into various convenient forms \cite{Gottesman_1997, Li_2008}.

For $k\neq0$, code equivalence together with the freedom to choose generators of the normalizer allows $G$ to be brought to the form 
\begin{equation} G = \begin{pmatrix} L & H^T \end{pmatrix}, \label{logical_stabilizer_form} \end{equation} 
where $L$ contains $2k$ independent logical Pauli operators and $H^T$ contains generators of the stabilizer group \cite{Gottesman_1997,Li_2008}. In this representation, the condition $\boldsymbol{x}\notin\mathcal{C}$ is equivalent to requiring that at least one of the first $2k$ components of $\boldsymbol{x}$ is nonzero. The self-dual and non-self-dual cases can be written uniformly by defining
\begin{equation} \mathcal{I} = \begin{cases} \{1,\ldots,2k\}, & k\neq0,\\ \{1,\ldots,n\}, & k=0, \end{cases} \qquad L_{\mathcal{I}} := |\mathcal{I}|, \label{constraint_set} \end{equation}
and 
\begin{equation} S(\boldsymbol{x}) := \sum_{i\in\mathcal{I}}x_i. \label{constraint_sum} \end{equation} The desired constraint is simply 
\begin{equation} S(\boldsymbol{x})\geq1. \label{nonzero_constraint} \end{equation} Introduce an auxiliary integer $h$ with \begin{equation} 1\leq h\leq L_{\mathcal{I}} \end{equation} and add the penalty 
\begin{equation} \Pi(\boldsymbol{x}) = \lambda \min_{1\leq h\leq L_{\mathcal{I}}} \left( S(\boldsymbol{x})-h \right)^2. \label{constraint_term} \end{equation}
If $S(\boldsymbol{x})\geq1$, choosing $h=S(\boldsymbol{x})$ makes the penalty vanish. If $S(\boldsymbol{x})=0$, the smallest possible value is obtained for $h=1$, and the penalty is $\lambda$. Thus, 
\begin{equation} \Pi(\boldsymbol{x}) = \begin{cases} 0, & S(\boldsymbol{x})\geq1,\\ \lambda, & S(\boldsymbol{x})=0. \end{cases} \label{penalty_values} \end{equation}
Since the Pauli weight cannot exceed $n$, choosing 
\begin{equation} \lambda=n+1 \label{penalty_weight} \end{equation}
ensures that every forbidden configuration has a larger total cost than any admissible configuration. The integer $h$ is represented by 
\begin{equation} h = 1+ \sum_{\ell=0}^{s_h-1} 2^\ell e_\ell, \qquad e_\ell\in\mathbb{Z}_2, \label{binary_h} \end{equation}
with 
\begin{equation} s_h = \left\lceil \log_2 L_{\mathcal{I}} \right\rceil. \label{define_sh} \end{equation} 

The constrained minimum-distance problem is therefore replaced by
\begin{equation}
d
=
\min_{\boldsymbol{x},\boldsymbol{y},\boldsymbol{e}}
\left[
F(\boldsymbol{x},\boldsymbol{y})
+
\Pi(\boldsymbol{x},\boldsymbol{e})
\right].
\label{constrained_min}
\end{equation}

The constraint introduces only $O(\log n)$ additional binary variables and does not change the overall $O(n\log n)$ scaling. The corresponding augmentation of the QUBO matrix is given in Appendix~\ref{app:assembly_penalty}. 

\subsubsection{Constraint as multiple cost functions} 
\label{sec:multiple_cost} 
The constraint can alternatively be enforced without introducing a penalty. For each $j=1,\ldots,L_{\mathcal{I}}$, consider the sector 
\begin{equation} \Omega_j = \left\{ \boldsymbol{x}\;:\; x_1=\cdots=x_{j-1}=0,\; x_j=1 \right\}. \label{constraint_sectors} \end{equation}
Every configuration satisfying Eq.~\eqref{nonzero_constraint} belongs to exactly one such sector, according to the position of its first nonzero component. Therefore, 
\begin{equation}
d
=
\min_{1\leq j\leq L_{\mathcal{I}}}
\left[
\min_{\substack{
\boldsymbol{x}\in\Omega_j\\
\boldsymbol{y}
}}
F(\boldsymbol{x},\boldsymbol{y})
\right].
\label{multiple_qubo}
\end{equation}

The original constrained problem can thus be replaced by $L_{\mathcal{I}}=O(n)$ unconstrained QUBOs, with some of the original variables fixed in each one. Although this generally requires more QUBO solves than the penalty approach, it can be advantageous when symmetries make the different sectors equivalent. 
\paragraph{Circulant self-dual codes.} 
\label{sec:circulant}
Consider a self-dual code with 
\begin{equation} G^T = \begin{pmatrix} I_n & B \end{pmatrix}, \end{equation}
where $B$ is circulant. A cyclic permutation of the components of $\boldsymbol{x}$ induces the same cyclic permutation of $B\boldsymbol{x}$ and therefore leaves the Pauli weight invariant. Any nonzero configuration can consequently be cyclically shifted so that a nonzero component occupies the first position. The constraint $\boldsymbol{x}\neq\boldsymbol{0}$ can therefore be imposed simply by fixing 
\begin{equation} x_1=1. \label{circulant_constraint} \end{equation}
Hence, for circulant self-dual codes, only a single unconstrained QUBO needs to be solved.

\section{Adiabatic Quantum Computing and Quantum Annealing}

Adiabatic Quantum Computing (AQC) is a universal model for quantum computation based on the adiabatic theorem. AQC starts by initializing the quantum processor in an easy-to-prepare ground state of some chosen initial Hamiltonian, $H(0) = H_I$. Subsequently, the system's Hamiltonian is slowly evolved over an extended time, $t_a$, transitioning towards a problem Hamiltonian, $H(t_a)=H_p$, whose ground state encodes the solution to a problem of interest. The adiabatic theorem guarantees that the evolved state $\ket{\Psi(t_a)}$ will have a large overlap with the ground state of the problem Hamiltonian provided  that the evolution was done slow enough \cite{farhi2000quantum}. In fact, this overlap, often termed the probability of success, $P_s := \abs{\braket{0,t_a}{\Psi(t_a)}}^2 $ satisfies
\begin{equation}
    \lim\limits_{t_a \to \infty} P_s = 1.
\end{equation}
Here, the state $\ket{m,t}$ satisfies $H(t)\ket{m,t} = E_m(t) \ket{m,t}$ with $E_m(t)$ denoting the $m$-th lowest eigenvalue of $H(t)$ . 

For Hamiltonians that can be written as $H(t)=\Tilde{H}(\gamma)$, where $\gamma:=t/t_a \in [0,1]$, and $\Tilde{H}$ has no explicit dependence on $t_a$,  the adiabatic condition can be expressed as
\begin{equation} 
\begin{aligned}
    t_a \gg&   \frac{\hbar \ \mathcal{E}}{g^2_{\text{min}}}, \\  &\mathcal{E}:= \max\limits_{\gamma} \abs{\left\langle 1,\gamma \middle| \partial_{\gamma} \Tilde{H} \middle| 0,\gamma \right\rangle},  \\ &g_{\text{min}} := \min\limits_{\gamma} \abs{E_1(\gamma) - E_0(\gamma)}.
    \end{aligned}
    \end{equation}
This condition ensures that $P_s$ is finite for finite values of $t_a$ \cite{farhi2000quantum}.  The performance of AQC depends on the chosen adiabatic path, specifically the choice of the initial Hamiltonian and the trajectory of evolution towards the problem Hamiltonian. For Hamilotnians exhibiting final  ground state degeneracy, $g_{\text{min}}$ is calculated between the instantaneous ground state and the first excited state that does not evolve eventually into the final ground state subspace \cite{Dickson_2011}. In this scenario,  we define the probability of success, $P_s$ , as the overlap between the final evolved state and the subspace of final degenerate ground-states.  Notably, AQC has been shown to be polynomially equivalent, in terms of the resource overhead, to universal gate-based quantum computation \cite{aharonov2005adiabatic}. 

Quantum Annealing (QA) is a heuristic approach that follows the same process as AQC but relaxes the condition of  adiabaticity. The departure from adiabatic evolution introduces excitations in the system's state, voiding the guarantee of convergence to the ground state of the problem Hamiltonian. To mitigate the limitations of nonadiabaticity, the algorithm repeats the entire evolution multiple times, generating various candidate final states. The algorithm then reports the best overall result. The computational power of QA and whether it offers a speedup over classical computation is currently an open area of research \cite{AdiabaticQuantumComputingandQuantumAnnealing}, with studies comparing the performance of QA  against classical algorithms like Simulated Annealing (SA) and Simulated Quantum Annealing (SQA) \cite{Crosson_2016}.

Quantum annealers are devices designed to run the QA algorithm. D-Wave systems have introduced multiple commercially available quantum annealers \cite{DWave}. In this paper, we focus on the D-wave Advantage 4.1 quantum annealer, which operates by evolving a mixed-field quantum Ising Hamiltonian of the form
\begin{equation}
\begin{aligned}
    H(\gamma) =& \underbrace{-\frac{\mathcal{A}(\gamma)}{2} \sum_i X_{(i)}}_\text{initial Hamiltonian} \\
    &+ \underbrace{\frac{\mathcal{B}(\gamma)}{2} \left( \sum_i h_i Z_{(i)} + \sum_{i>j} J_{ij} Z_{(i)} Z_{(j)}\right)}_\text{problem Hamiltonian}.
    \end{aligned}
\end{equation}
Here, $\mathcal{A}(\gamma)$ and $\mathcal{B}(\gamma)$ are functions that control the transition between the initial and problem Hamiltonians, satisfying $\mathcal{A}(0)\gg \mathcal{B}(0)$ and $\mathcal{A}(1)\ll \mathcal{B}(1)$. The parameters $\gamma$ and $t_a$ are referred to as the annealing parameter and the annealing time, respectively. The existence of a map between any QUBO cost function and the problem Hamiltonian of the quantum Ising model \cite{Lucas_2014} makes the D-Wave Advantage 4.1 quantum annealer a natural tool to solve the QUBOs constructed in the previous section.  The exact functional forms of $\mathcal{A}(\gamma)$ and $\mathcal{B}(\gamma)$ as implemented on the device, can be found in \cite{DWave}. More information about the device is provided in the appendix.

\subsection{Necessary runtime analysis}
The  runtime necessary to evaluate the optimum for the worst-case instance of an NP-complete optimization problem is  believed to be exponentially large in problem size for both classical and quantum  algorithms. Hypothetically however, it could be that quantum algorithms offer a modest (sub-exponential) speedup. It is also conceivable that quantum algorithms may offer systematically better performance on average, i.e.~for typical instances \cite{Albash_2018}. In practice there are many factors affecting   AQC runtime necessary to solve a  QUBO problem e.g.~associated with the circulant self-dual optimal stabilizer codes. 
We discuss the most important of them below, assuming for simplicity
that the quantum device possesses sufficient connectivity to enable direct mapping of QUBO variables onto the physical qubits. The effects of limited connectivity, which are by far the most important, will be discussed in later sections.

\begin{figure*}[!htb]
\centerline{\includegraphics[width=\textwidth]{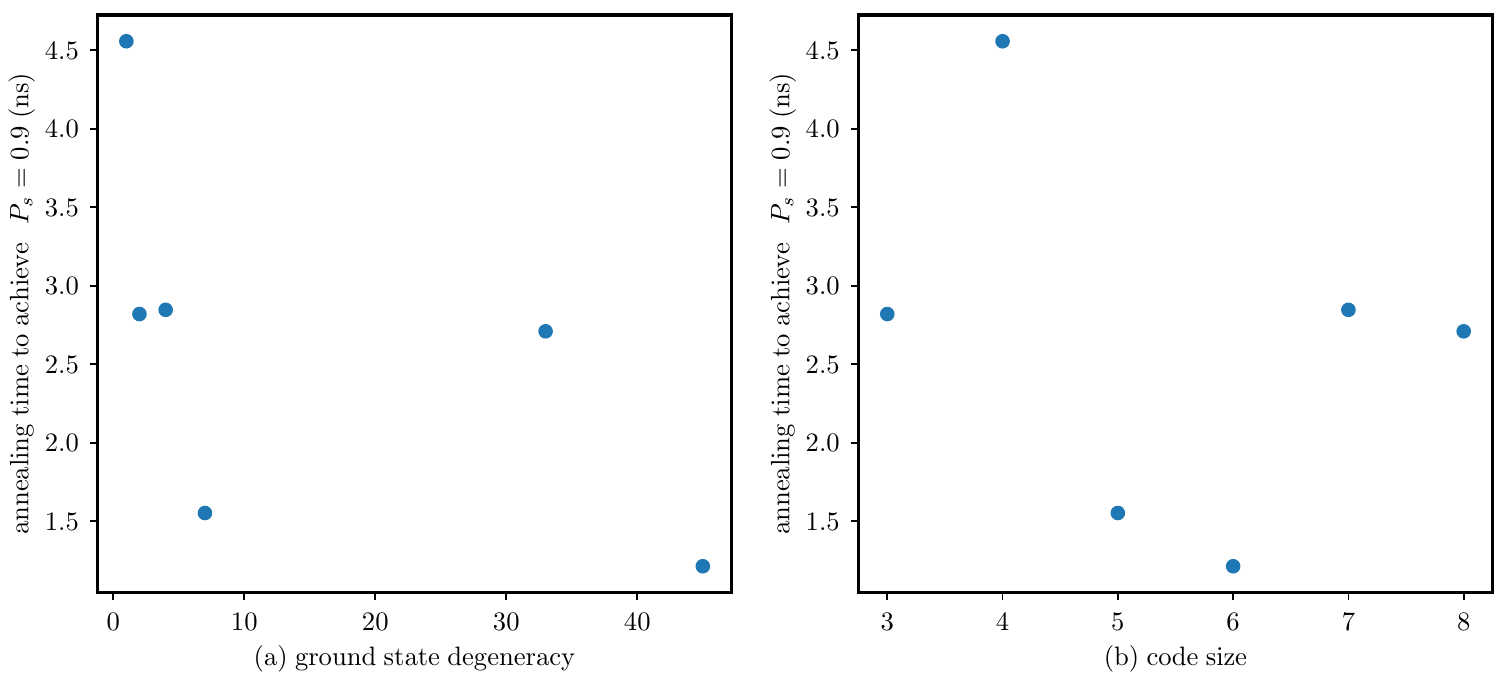}}
\caption{\textbf{Estimation of necessary annealing time.} Both panels show the annealing time required to achieve $P_s = 0.9$ ($P_s$ stands for probability of success), calculated using the form of $\mathcal{A}(s)$ and $\mathcal{B}(s)$ utilized by the D-Wave Advantage 4.1 quantum annealer. a) depicts the relationship between this annealing time and the ground-state degeneracy of the problem Hamiltonian. b) shows the same annealing time against the code size. For larger code sizes, the dimensions of the corresponding Ising Hamiltonian exceeded  computational feasibility.}
    \label{probsucc}
\end{figure*}

It is well-known that the degeneracy of the ground state, even in absence of quantum noise and decoherence, may significantly influence necessary annealing time $t_a$.
Given that circulant codes are cyclic, our problem Hamiltonian  $H(\gamma=1)$ always exhibits multiple ground states. Ground state degeneracy can both increase and decrease $t_a$, depending on the degeneracy of the excited states \cite{Matsuda_2009,King_2016,Dickson_2011}. To address this question in our case we numerically simulated the time-dependent Schrodinger equation describing the annealing process for codes of sizes $n=3-8$. In these simulations annealing time was varied until the success rate $P_s= 0.9$ was achieved. The results, shown in the panel a) of Fig.~1 are consistent with the inverse correlation between the ground state degeneracy and minimal necessary $t_a$. 

This conclusion is very preliminary, as the degeneracy of final Hamiltonian is merely one of the factors. Another important factor is the smallest value of gap of $H(\gamma)$, which may change significantly for large $n$ (we checked that for codes of sizes $n=3-7$ it does not). For the Hamiltonian employed, it can be demonstrated that $\mathcal{E}$ scales polynomially with the problem size, as follows from the observation that the eigenvalues of $H_I$ and $H_P$ are also at most polynomial. Yet,  understanding the scaling of $g_{\text{min}}$ with $n$ is usually a very difficult problem, see \cite{Albash_2018}.

Our classical simulations of the annealing dynamics, shown in Fig. 1, panel b), suggest that theoretically required  annealing times for the small code sizes $n\sim 3-8$  fall within the nanosecond range. In practice, as we discussed in the next section, reasonable necessary annealing time on the D-Wave Advantage 4.1 platform for these problems is of the order of microseconds. This suggests 
theoretically quantum annealer is operating firmly within the adiabatic regime, but in practice its quality of performance is dominated by other factors related to quantum noise. The factors discussed above are much less important, at least for moderate problem sizes $n$. This instructs our approach in the next section to optimize the annealing time $t_a$ empirically.

\section{Results of Quantum Demonstrations on the D-Wave QA Platform}

\subsection{Quantum-only implementation}

\begin{figure}[!htb]
\centerline{\includegraphics[width=\columnwidth]{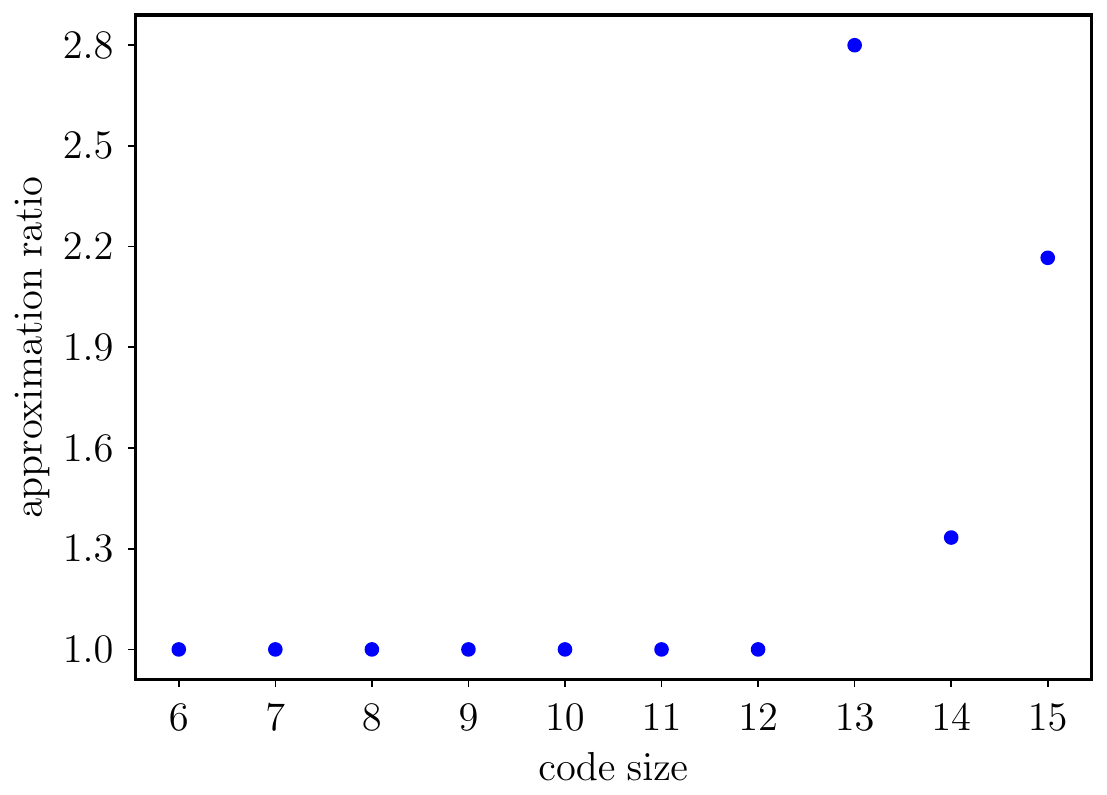}}
\caption{\textbf{Quantum Annealing performance for small code size.} This figure plots the Approximation ratio (AR) of best output generated by runs of quantum annealer solving the QUBO associated with optimal circulant codes of length $6\leq n\leq 15$. AR equals one means the algorithm found the global minimum (exact value of the minimum code distance). Larger AR means that despite numerous runs, the algorithm failed to find the global minimum and only produced a conservative bound.  We performed 40 runs (one run is a 100 anneals) for each code size.\\
}
\label{fig:approx_ratio}
\end{figure}

We tested our QUBO-based algorithm by running implementations on the D-Wave Advantage 4.1 quantum annealer. We ran instances of optimal circulant self-dual codes of small length, taken from the database composed by M.~Grassl \cite{Grassl:codetables}. The constraint $\boldsymbol{x} \neq 0^n$  was imposed by setting $x_1 =1$ as described in Sec.~\ref{sec:circulant}. To quantify how close the quantum annealer got to the optimal value, we calculated the so-called approximation ratio for various code sizes. The approximation ratio is the ratio between the smallest value generated by the quantum annealer and the actual minimum distance. As shown in Figure~\ref{fig:approx_ratio}, the quantum annealer was able to successfully find exact minimum distance for codes of sizes up to $12$. For larger code sizes, despite numerous re-runs, QA could only find conservative upper bounds. Our results strongly suggest that QA alone can not currently compete with purely classical algorithms/implementations, but there could be a room to use QA  as a tool to obtain rigorous upper bounds. 

To systematically analyze the performance of quantum annealer, 
we study how the quality of outcome depends on the annealing time $t_a$,
a free parameter of implementation of the quantum algorithm. At least naively, longer annealing time would make time evolution more adiabatic. But since the platform is noisy, a longer annealing time can lead to an accumulation of errors. Thus, we can expect that certain values of $t_a$ would yield optimal performance.  

\begin{figure}[!htb]
\centerline{\includegraphics[width=\columnwidth]{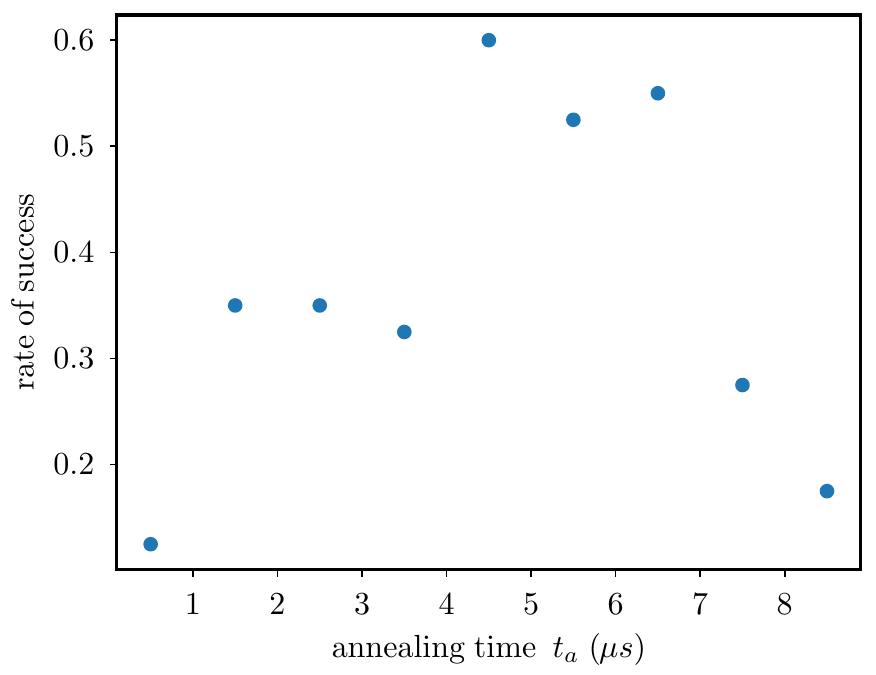}}
\caption{\textbf{Effects of increasing annealing time.} This figure depicts the success rate of the QUBO-based algorithm run on D-Wave Advantage 4.1 as a function of different annealing times $t_a$. Each data point represents the success rate, calculated as the number of times the optimal value was achieved out of 40 runs (each comprising a 100 anneals). All runs were optimizing the same QUBO associated with an optimal circulant code of length $n=9$. The annealing times ranged from $0.5$ to $8.5  \mu s$ in increments of $1 \mu s$.}
\label{fig:code9}
\end{figure}

To study $t_a$ dependence, we plot the rate of success for the QUBO associated with the given code of length $n=9$ using various annealing times. The rate of success is defined as the proportion of runs where the optimal result was obtained out of the total number of runs conducted. As shown in Figure~\ref{fig:code9}, the rate of success initially increases as we increase the annealing time, reaching its peak at approximately $4.5 \ \mu s$. However, increasing the annealing time beyond this point resulted in deteriorating performance. Considering the results from the previous section, which suggested that, at least theoretically, we are operating within the adiabatic regime, this deterioration in performance is likely due to accumulated noise.  Empirically, we find that an annealing time of $t_a = 4.5 \ \mu s$ is close to optimum for most code sizes, and therefore, we  use this value in all of our quantum demonstrations.

\begin{figure}[!htb]
\centerline{\includegraphics[width=\columnwidth]{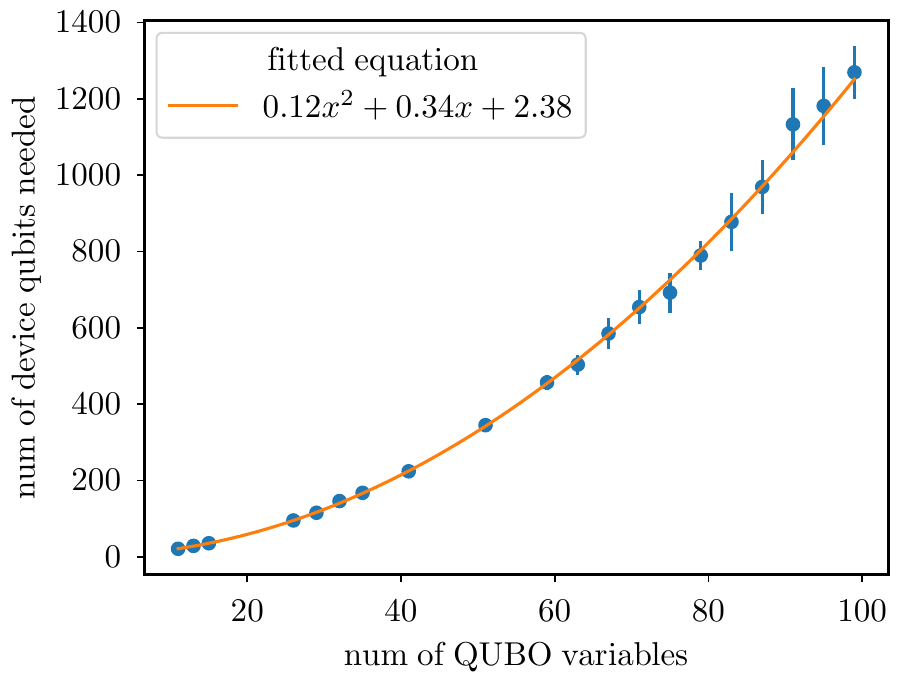}}
\caption{\textbf{Qubit overhead needed to satisfy connectivity constraints.} This figure shows the mean number of physical qubits employed by the QA platform (i.e.~number of qubits after the so-called embedding) vs.~the number of binary variables of the underlying QUBO problem. 
Standard deviation, due to heuristic nature of the embedding algorithm, is shown as the error bars. It is usually small. 
To generate the plot, we used 
optimal circulant codes of different sizes, reformulated the minimum distance problem as QUBO and then embedded it onto the QA platform 10 times. We note that QUBO associated with different codes can have different  connectivity requirements, which can affect the number of qubits after embedding. 
The data points (blue circles) conform well to quadratic fit (orange curve), which suggests that on average the number of qubits after embedding grows quadratically.    }
\label{fig:connectivity overhead}
\end{figure}

Suboptimal performance of quantum annealer for relatively small code sizes $n\geq 13$ can likely be attributed to  limited connectivity between physical qubits of the platform. On the annealer, a quadratic interaction between two QUBO variables requires
a programmable coupling between their corresponding physical qubits.
Because the connectivity of the hardware graph is limited, not every pair of
interacting logical variables can be represented directly. Minor embedding
addresses this by representing a single logical variable by a chain of
physical qubits \cite{cai2014practical}, thereby enlarging the physical
optimization problem relative to the logical QUBO. This leads to an overhead in the number of qubits employed by the platform. Figure~\ref{fig:connectivity overhead} shows that the number of necessary physical qubits increases approximately {\it quadratically} with the size of the original QUBO problem, associated with the code. The increase in the number of necessary qubits significantly complicates successful implementation of the QA algorithm. As the number of physical qubits grows, the corresponding search space also expands. Furthermore, since current quantum devices are noisy, each additional qubit introduces additional sources of noise that can affect the final result of the algorithm.

\subsection{Hybrid implementation}

To mitigate challenges of limited connectivity, we shift our focus to hybrid quantum-classical implementation. Specifically, we use Qbsolv algorithm, a hybrid algorithm developed by the D-Wave Systems. Qbsolv employs a classical algorithm to partition QUBO problems into smaller and more manageable pieces. These ``subproblems''  can be solved using QA, or any other suitable QUBO solving algorithm, before stitching them together \cite{Booth2017PartitioningOP}. This approach offers two significant benefits. First, it allows solving problems of bigger size, exceeding current limits of  quantum platforms. Second, the subproblems have fewer connectivity constraints, which reduces the number of necessary physical qubits,  and in this way decreases the source of errors. 

To compare the hybrid QA-assisted implementation with a fully classical one,
we solved the Qbsolv subproblems using either quantum annealing on the D-Wave
platform or classical simulated annealing (SA). We refer to these as the
``QA-assisted'' and ``SA-assisted'' implementations, respectively. At each
code length, the two implementations were assigned the same end-user time
budget and the same number of Qbsolv runs, although these quantities varied
with the code size $n$.

This is a comparison of the two implementations from the perspective of a
user of the current platforms, rather than a controlled comparison of two
solvers performing identical computational work. In particular,  the binary encodings supplied to the two implementations were not always identical in size, with the QA-assisted
instances generally being larger. In addition, the QA implementation incurs
the minor-embedding overhead described above, so that its logical QUBO is
represented by a substantially larger physical Ising problem. Remote execution of QA on the D-Wave platform also introduces latency. These effects all act against the present QA implementation and should be distinguished from the
intrinsic difficulty of the underlying logical QUBO. Since the subproblems are of similar size to those considered in the previous
subsection, we use an annealing time of $t_a=4.5\,\mu\mathrm{s}$ for the
QA-assisted demonstrations.

The two QUBO-based implementations show broadly comparable performance.
Across the 31 code lengths retained in Fig.~\ref{fig:qbsolv}, the QA-assisted
implementation reaches the tabulated minimum distance at 18 lengths and the
SA-assisted implementation at 17. Neither implementation consistently
dominates the other, and both return progressively looser upper bounds for
some of the larger instances. Figure~\ref{fig:qbsolv} also places these results alongside specialized
classical distance-finding methods, which we discuss in the following
subsection.

\subsection{Comparison with classical minimum distance algorithms}
\label{sec:classical-comparison}

To place the QUBO formulation in context, we performed a direct numerical
comparison with specialized classical distance-finding methods. We used
BZDistMW as a Brouwer--Zimmermann exact method and QDistRndMW as
a heuristic method, using the implementations in the \texttt{codeDistance}
package (version 0.0.8) \cite{Webster_2026}. Brouwer--Zimmermann certifies the
minimum distance when allowed to terminate, whereas the heuristic methods
return upper bounds that need not in general be optimal.

For a present-day end-user comparison, we imposed on each classical method a
wall-clock cap equal to the total budget of the QUBO runs at the same code length. This gives a meaningful comparison of what the
available implementations return within the same user-visible time allowance,
but it should not be interpreted as a hardware-independent measure of
computational work. Classical processors and software are highly mature,
whereas the present QA implementation incurs additional architectural costs, most notably minor embedding of the logical QUBO into a substantially larger
physical Ising problem.

QDistRnd returned the tabulated minimum distance at all tested lengths within the imposed budget. In contrast, the QA- and SA-assisted QUBO implementations reach the
tabulated minimum distance at 18 and 17 of the 31 lengths, respectively.
Thus, on the instances studied here, the present QUBO implementations do not
provide a computational advantage over specialized classical heuristic
distance-finding methods.

\begin{figure*}[!htb]
\includegraphics[width=\textwidth]{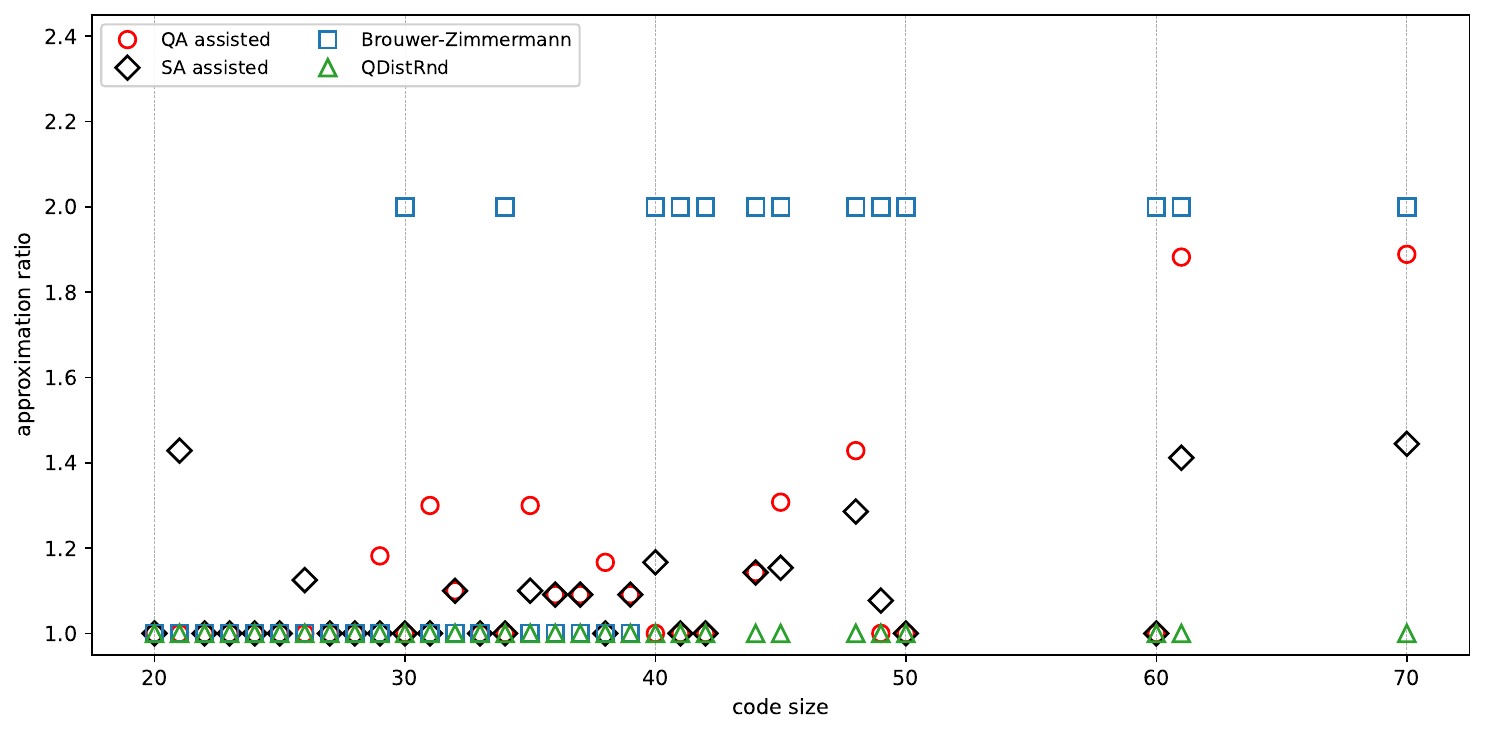}
\caption{\textbf{Comparison of QUBO-based and specialized classical distance-finding methods.}
The approximation ratio is the returned upper bound divided by the tabulated
minimum distance. Red circles and black diamonds show the QA-assisted and SA-assisted QUBO results, respectively. Blue squares show
Brouwer--Zimmermann runs with an enforced wall-clock cap equal to the total time budget of the corresponding QUBO runs, while green triangles
show QDistRnd under the same cap. The figure compares results obtained under matched end-user wall-clock budgets, rather than equal computational
work across hardware platforms.}
\label{fig:qbsolv}
\end{figure*}

The exact Brouwer--Zimmermann comparison illustrates a different tradeoff.
In the capped sweep shown in Fig.~\ref{fig:qbsolv}, Brouwer--Zimmermann
reaches the tabulated distance at 18 of the 31 tested lengths and returns an
approximation ratio of $2$ at the remaining 13. When allowed
to complete, Brouwer--Zimmermann certified the tabulated minimum distance at
all tested lengths from $n=20$ through $n=40$.

\section{Discussion}

In this work we introduced a reformulation of the minimum-distance problem for
binary quantum stabilizer codes as a quadratic unconstrained binary
optimization problem. The formulation uses $O(n\log n)$ binary variables and
can be supplied without modification to classical, quantum, or hybrid QUBO
solvers. This solver-independent representation is the main conceptual
advantage of the approach and can benefit directly from future developments
in general-purpose QUBO optimization.

We tested the formulation using the D-Wave Advantage 4.1 quantum annealer and
the hybrid Qbsolv framework. The purely quantum implementation ceases to
reliably reach the optimum at relatively small code lengths, while the
QA-assisted and SA-assisted Qbsolv implementations perform substantially
better and are broadly comparable to one another. The direct classical
benchmark of Sec.~\ref{sec:classical-comparison}, however, shows that
specialized classical distance-finding methods remain more effective on the
instances considered here. In particular, QDistRnd return the
tabulated minimum distance throughout the tested data set under the same
end-user wall-clock budgets. We therefore find no evidence of a computational
advantage for the present quantum-annealing implementation.

The wall-clock comparison answers a practical present-day question: given the
currently available implementations and hardware, what result can an end user
obtain within a fixed amount of time? It should not be interpreted as a
hardware-neutral comparison of intrinsic computational work; the present work
does not attempt such a comparison. Classical processors and  optimization software are highly mature, whereas
current quantum-annealing hardware is at a much earlier stage and incurs
substantial implementation overhead. In particular, limited connectivity
requires minor embedding, so the annealer operates on a physical Ising problem
that can be considerably larger than the logical QUBO. Improvements in
connectivity and hardware quality could reduce this disadvantage in future
generations.

For the current Advantage 4.1 platform, the dominant hardware limitation is
minor embedding rather than the logical variable count alone. The QUBO
formulation requires only $O(n\log n)$ logical binary variables, whereas
Fig.~\ref{fig:connectivity overhead} shows that the number of physical qubits
required after embedding grows approximately quadratically with the number of
QUBO variables. Combining these observed scalings gives an effective physical
qubit requirement of roughly $O(n^2\log^2 n)$ for the present architecture. Higher
connectivity would reduce the embedding overhead directly, while lower noise
would make the resulting physical embeddings more reliable.

Although the present implementation does not outperform specialized classical
distance-finding methods, the QUBO reformulation remains applicable to other
optimization platforms. Future work could explore improved classical QUBO
solvers, later generations of quantum annealers, or gate-based approaches such
as the Quantum Approximate Optimization Algorithm (QAOA)
\cite{farhi2014quantum}.

To conclude, we proposed a reformulation of a theoretically and practically important NP-hard problem as a QUBO problem, contributing to an ongoing effort to reformulate various problems in a form suitable for quantum platforms. Looking ahead, an important step would be to extend our approach to other problems of classical and quantum information theory and discrete mathematics. In particular, the shortest lattice vector problem, a known NP-hard problem, which is conceptually similar to code minimum distance problem discussed in this work, would be a natural  next candidate. 

\section{Acknowledgments}
We thank D-Wave for providing access to their quantum cloud services, Leap. This research is supported by the NSF grant PHY 2310426. 

\bibliographystyle{utphys}
\bibliography{references}

\appendix

\section{Assembly of the QUBO matrices}
\label{app:assembly}

This appendix gives the explicit QUBO matrices corresponding to the
reformulations introduced in Sec.~\ref{sec:algorithm}. We use throughout the
notation introduced in that section.

\subsection{QUBO matrix for the Pauli weight}
\label{app:assembly_general}

For each $r=1,2,3$, collect the binary variables $q_{i\ell}^{(r)}$ into
\begin{equation}
\boldsymbol{q}^{(r)}
=
\begin{pmatrix}
q_{10}^{(r)} &
\cdots &
q_{n0}^{(r)} &
q_{11}^{(r)} &
\cdots &
q_{n,s_r-1}^{(r)}
\end{pmatrix}^{T},
\end{equation}
and define
\begin{equation}
D_r
:=
\begin{pmatrix}
1 & 2 & \cdots & 2^{s_r-1}
\end{pmatrix}
\otimes I_n.
\label{define_Dr}
\end{equation}
Then the auxiliary integers in Eq.~\eqref{binary_q} can be written collectively as
\begin{equation}
\begin{pmatrix}
q_1^{(r)}\\
\vdots\\
q_n^{(r)}
\end{pmatrix}
=
D_r\boldsymbol{q}^{(r)}.
\label{binary_q_vector}
\end{equation}

Using
\begin{equation}
\boldsymbol{m}^{(r)}
=
C^{(r)}\boldsymbol{x},
\end{equation}
Eq.~\eqref{integer_quadratic} becomes
\begin{equation}
F
=
\frac{1}{2}
\sum_{r=1}^{3}
\left\|
C^{(r)}\boldsymbol{x}
-
2D_r\boldsymbol{q}^{(r)}
\right\|_2^2.
\label{compact_qubo}
\end{equation}
Each term expands as
\begin{equation}
\begin{aligned}
\frac{1}{2}
\left\|
C^{(r)}\boldsymbol{x}
-
2D_r\boldsymbol{q}^{(r)}
\right\|_2^2
={}&
\frac{1}{2}
\boldsymbol{x}^{T}
\left(C^{(r)}\right)^T
C^{(r)}
\boldsymbol{x}
\\
&-
2\boldsymbol{x}^{T}
\left(C^{(r)}\right)^T
D_r
\boldsymbol{q}^{(r)}
\\
&+
2
\left(\boldsymbol{q}^{(r)}\right)^T
D_r^T D_r
\boldsymbol{q}^{(r)}.
\end{aligned}
\label{generic_expansion}
\end{equation}

For later convenience, collect the three families of parity auxiliary
variables into
\begin{equation}
\boldsymbol{y}
=
\begin{pmatrix}
\boldsymbol{q}^{(1)}\\
\boldsymbol{q}^{(2)}\\
\boldsymbol{q}^{(3)}
\end{pmatrix}.
\label{parity_aux_vector}
\end{equation}
The variables of the weight QUBO are therefore
\begin{equation}
\begin{pmatrix}
\boldsymbol{x}\\
\boldsymbol{y}
\end{pmatrix},
\end{equation}
and
\begin{equation}
F(\boldsymbol{x},\boldsymbol{y})
=
\begin{pmatrix}
\boldsymbol{x}\\
\boldsymbol{y}
\end{pmatrix}^{T}
Q
\begin{pmatrix}
\boldsymbol{x}\\
\boldsymbol{y}
\end{pmatrix},
\end{equation}

with the symmetric QUBO matrix. To keep its block structure compact, define
\begin{equation}
R_r
:=
-\left(C^{(r)}\right)^T D_r,
\qquad
S_r
:=
2D_r^T D_r,
\qquad
r=1,2,3.
\label{define_RS}
\end{equation}
Then
\begin{equation}
Q
=
\begin{pmatrix}
Q_{xx} & R_1 & R_2 & R_3 \\
R_1^T & S_1 & 0 & 0 \\
R_2^T & 0 & S_2 & 0 \\
R_3^T & 0 & 0 & S_3
\end{pmatrix},
\label{matrix_eq}
\end{equation}
where
\begin{equation}
Q_{xx}
=
\frac{1}{2}
\sum_{r=1}^{3}
\left(C^{(r)}\right)^T C^{(r)}.
\label{Qxx}
\end{equation}

Using Eq.~\eqref{define_Cr},
\begin{equation}
Q_{xx}
=
A^TA+B^TB
+
\frac{1}{2}
\left(
A^TB+B^TA
\right).
\label{Qxx_AB}
\end{equation}

\subsection{QUBO matrix for self-dual codes}
\label{app:assembly_selfdual}

For the graph-code representation of Eq.~\eqref{graph_code_form}, collect the
auxiliary binary variables into $\boldsymbol{q}$ and define
\begin{equation}
D
:=
\begin{pmatrix}
1 & 2 & \cdots & 2^{s_B-1}
\end{pmatrix}
\otimes I_n,
\end{equation}
so that the corresponding auxiliary integers are given collectively by
$D\boldsymbol{q}$.  Equation~\eqref{selfdual_quadratic} can then be written as
\begin{equation}
F_{\mathrm{SD}}
=
\boldsymbol{x}^{T}\boldsymbol{x}
+
\left(
B\boldsymbol{x}
-
2D\boldsymbol{q}
\right)^T
\left(
B\boldsymbol{x}
-
2D\boldsymbol{q}
-
\boldsymbol{x}
\right).
\label{compact_selfdual}
\end{equation}
Since $B$ is the adjacency matrix of an undirected graph,
\begin{equation}
B=B^T.
\end{equation}
Expanding Eq.~\eqref{compact_selfdual} therefore gives
\begin{equation}
F_{\mathrm{SD}}
=
\begin{pmatrix}
\boldsymbol{x}\\
\boldsymbol{q}
\end{pmatrix}^{T}
Q_{\mathrm{SD}}
\begin{pmatrix}
\boldsymbol{x}\\
\boldsymbol{q}
\end{pmatrix},
\end{equation}
where
\begin{equation}
Q_{\mathrm{SD}}
=
\begin{pmatrix}
I_n+B^2-B
&
(I_n-2B)D
\\
D^T(I_n-2B)
&
4D^TD
\end{pmatrix}.
\label{matrix_selfdual}
\end{equation}

\subsection{QUBO matrix for the constraint penalty}
\label{app:assembly_penalty}

We now derive the modification of the QUBO matrix produced by the
non-triviality constraint. Recall that the constraint requires at least one
component of $\boldsymbol{x}$ in the relevant set $\mathcal{I}$ to be nonzero,
where
\begin{equation}
\mathcal{I}
=
\begin{cases}
\{1,\ldots,2k\}, & k\neq 0,\\
\{1,\ldots,n\}, & k=0.
\end{cases}
\end{equation}
Thus,
\begin{equation}
\sum_{i\in\mathcal{I}}x_i \geq 1.
\label{app_nonzero_constraint}
\end{equation}

As described in Sec.~\ref{sec:constraint_penalty}, this condition is enforced
by adding the penalty
\begin{equation}
\Pi
=
\lambda
\min_{1\leq h\leq |\mathcal{I}|}
\left(
\sum_{i\in\mathcal{I}}x_i-h
\right)^2.
\label{app_constraint_penalty}
\end{equation}
The auxiliary integer $h$ is represented by
\begin{equation}
h
=
1+\sum_{\ell=0}^{s_h-1}2^\ell e_\ell,
\qquad
e_\ell\in\mathbb{Z}_2,
\label{app_binary_h}
\end{equation}
with
\begin{equation}
s_h
=
\left\lceil
\log_2|\mathcal{I}|
\right\rceil.
\end{equation}

To write Eq.~\eqref{app_constraint_penalty} compactly, define the vector
$\boldsymbol{v}$ by
\begin{equation}
v_i
=
\begin{cases}
1, & i\in\mathcal{I},\\
0, & i\notin\mathcal{I},
\end{cases}
\label{constraint_indicator}
\end{equation}
so that
\begin{equation}
\sum_{i\in\mathcal{I}}x_i
=
\boldsymbol{v}^T\boldsymbol{x}.
\end{equation}
Likewise, collect the binary variables $e_\ell$ into
\begin{equation}
\boldsymbol{e}
=
\begin{pmatrix}
e_0 & e_1 & \cdots & e_{s_h-1}
\end{pmatrix}^T,
\end{equation}
and define
\begin{equation}
\boldsymbol{w}
=
\begin{pmatrix}
1 & 2 & \cdots & 2^{s_h-1}
\end{pmatrix}^T.
\label{constraint_weights}
\end{equation}
Then
\begin{equation}
h
=
1+\boldsymbol{w}^T\boldsymbol{e},
\end{equation}
and the penalty becomes
\begin{equation}
\Pi
=
\lambda
\left(
\boldsymbol{v}^T\boldsymbol{x}
-
1
-
\boldsymbol{w}^T\boldsymbol{e}
\right)^2.
\label{penalty_compact}
\end{equation}

Expanding the square gives
\begin{equation}
\begin{aligned}
\Pi
=
\lambda\Big[
&
(\boldsymbol{v}^T\boldsymbol{x})^2
-2\boldsymbol{v}^T\boldsymbol{x}
-2(\boldsymbol{v}^T\boldsymbol{x})
  (\boldsymbol{w}^T\boldsymbol{e})
\\
&
+(\boldsymbol{w}^T\boldsymbol{e})^2
+2\boldsymbol{w}^T\boldsymbol{e}
+1
\Big].
\end{aligned}
\label{penalty_expanded}
\end{equation}

Because the variables are binary, $x_i^2=x_i$ and $e_\ell^2=e_\ell$,
so the linear terms in Eq.~\eqref{penalty_expanded} can be placed on the
diagonal of the corresponding QUBO blocks. The three nonzero blocks
introduced by the constraint are therefore
\begin{equation}
\Delta Q_{xx}
=
\lambda
\left[
\boldsymbol{v}\boldsymbol{v}^T
-
2\,\mathrm{diag}(\boldsymbol{v})
\right],
\label{delta_Qxx}
\end{equation}
\begin{equation}
\Delta Q_{xe}
=
-\lambda\boldsymbol{v}\boldsymbol{w}^T,
\label{delta_Qxe}
\end{equation}
and
\begin{equation}
\Delta Q_{ee}
=
\lambda
\left[
\boldsymbol{w}\boldsymbol{w}^T
+
2\,\mathrm{diag}(\boldsymbol{w})
\right].
\label{delta_Qee}
\end{equation}
Here $\mathrm{diag}(\boldsymbol{v})$ denotes the diagonal matrix whose
diagonal entries are the components of $\boldsymbol{v}$, and similarly for
$\mathrm{diag}(\boldsymbol{w})$.

To add these terms to the weight QUBO, write the matrix $Q$ derived in
Appendix~\ref{app:assembly_general} in the block form
\begin{equation}
Q
=
\begin{pmatrix}
Q_{xx} & Q_{xy}\\
Q_{yx} & Q_{yy}
\end{pmatrix},
\label{Q_partition}
\end{equation}
where $Q_{xx}$ contains terms involving the original variables
$\boldsymbol{x}$, $Q_{yy}$ contains terms involving the parity auxiliary
variables $\boldsymbol{y}$, and $Q_{xy}=Q_{yx}^T$ contains their couplings.

After including the new constraint variables $\boldsymbol{e}$, the complete
variable vector is
\begin{equation}
\boldsymbol{z}
=
\begin{pmatrix}
\boldsymbol{x}\\
\boldsymbol{y}\\
\boldsymbol{e}
\end{pmatrix}.
\label{constrained_variable_vector}
\end{equation}
Since the constraint penalty depends only on $\boldsymbol{x}$ and
$\boldsymbol{e}$, it does not introduce any coupling between
$\boldsymbol{e}$ and the parity auxiliary variables $\boldsymbol{y}$.

The complete QUBO matrix then takes the compact block form
\begin{equation}
\mathcal{Q}
=
\begin{pmatrix}
Q_{xx}+\Delta Q_{xx} & Q_{xy} & \Delta Q_{xe} \\
Q_{yx} & Q_{yy} & 0 \\
\Delta Q_{xe}^T & 0 & \Delta Q_{ee}
\end{pmatrix}.
\label{matrix_constrained}
\end{equation}

The remaining constant term in Eq.~\eqref{penalty_expanded} is $\lambda$.
It does not affect the minimizing configuration, but it must be retained when
recovering the value of the minimum distance. Hence,
\begin{equation}
d
=
\lambda
+
\min_{\boldsymbol{z}}
\boldsymbol{z}^T\mathcal{Q}\boldsymbol{z}.
\label{constrained_min_matrix}
\end{equation}

\subsection{Example: a circulant self-dual code}
\label{app:assembly_example}

As a simple example, consider the length-$5$ circulant self-dual code with
\begin{equation}
G^T
=
\begin{pmatrix}
I_5 & B
\end{pmatrix},
\qquad
B
=
\operatorname{circ}(0,1,0,0,1).
\label{example_generator}
\end{equation}
The matrix $B$ is the adjacency matrix of the $5$-cycle. Since the code is in
the graph-code form of Eq.~\eqref{graph_code_form}, the reduced self-dual
construction of Sec.~\ref{sec:self_dual} applies.

Each row of $B$ contains two nonzero entries, so
\begin{equation}
M_B=2.
\end{equation}
The corresponding auxiliary integer satisfies
\begin{equation}
0\leq q_i\leq1,
\end{equation}
and therefore requires only one binary variable,
\begin{equation}
q_i=q_{i0},
\qquad
q_{i0}\in\mathbb{Z}_2.
\end{equation}
Thus the self-dual reformulation introduces five auxiliary binary variables,
one for each qubit.

For $s_B=1$, the matrix implementing the binary expansion is simply
\begin{equation}
D=I_5.
\end{equation}
Substituting this into Eq.~\eqref{matrix_selfdual} gives
\begin{equation}
Q_{\mathrm{SD}}
=
\begin{pmatrix}
I_5+B^2-B & I_5-2B \\
I_5-2B & 4I_5
\end{pmatrix}.
\label{example_Q}
\end{equation}
For the particular circulant matrix in Eq.~\eqref{example_generator},
\begin{equation}
I_5+B^2-B
=
\operatorname{circ}(3,-1,1,1,-1).
\end{equation}

It remains to impose the non-triviality condition
$\boldsymbol{x}\neq\boldsymbol{0}$. Because $B$ is circulant, a cyclic
permutation of $\boldsymbol{x}$ produces the same cyclic permutation of
$B\boldsymbol{x}$ and therefore leaves the Pauli weight unchanged. Every
nonzero configuration is consequently equivalent, under a cyclic shift, to
one with
\begin{equation}
x_1=1.
\end{equation}
The constraint can therefore be imposed by fixing $x_1$ rather than by introducing a penalty term as discussed in Sec.~\ref{sec:circulant}.

After fixing $x_1=1$, the resulting QUBO contains the remaining four
components of $\boldsymbol{x}$ together with the five auxiliary variables
$q_{i0}$, for a total of nine binary optimization variables.

\section{Quantum Hardware Characteristics}
\label{app:qpu-hardware}

All demonstrations in this work were executed on a D-Wave Advantage generation quantum annealer\cite{DWaveAdvantageOverview}, specifically the D-Wave Advantage 4.1 device ( referred to as ``Advantage\_system\_4.1'' in D-Wave's documentation \cite{DWaveAdvantageSystem4_1Properties}). The system was accessed through both Amazon Braket and the D-Wave Leap service. This device implements the Pegasus connectivity graph, which is the native topology for Advantage-generation QPUs. In the Pegasus topology, each qubit is connected to up to 15 neighboring qubits: 12 via internal  couplers and 3 via external or odd couplers. In practice, the realized connectivity in a given calibration cycle can be lower due to fabrication yield and calibration-related disabling of qubits and/or couplers. This Pegasus connectivity contrasts with newer D-Wave architectures, like Advantage2, which employ the Zephyr topology \cite{DWave}.

A Pegasus unit cell contains 24 qubits, and an Advantage QPU consists of a $16 \times 16$ lattice of such unit cells. The post-fabrication physical device used in our demonstrations reports 5627 qubits and 40279 couplers. For each QPU, an average qubit temperature is estimated from measurements of qubit population as a function of per-qubit linear bias as described in \cite{dwave-qpu-temperature}. For the D-Wave Advantage 4.1 system, the reported average temperature across qubits is $15.4 \pm 0.1~\mathrm{mK}$.

Table~\ref{tab:qpu-physical-properties} summarizes additional physical characteristics of the device that are relevant to quantum annealing performance, as reported for the D-Wave Advantage 4.1 system. Further details on the hardware architecture and qubit properties can be found in Refs.~\cite{DWave, DWaveAdvantageOverview, DWaveAdvantageSystem4_1Properties}.

\begin{table*}[!htb]
\centering

\begin{tabular}{ll}
\hline
\textbf{Property} & \textbf{Value} \\
\hline
Generation & Advantage \\
Topology & Pegasus \\
Name &  Advantage\_system4.1 \\
Number of qubits & 5627 \\
Number of couplers & 40279 \\
Average qubit temperature & $15.4 \pm 0.1~\mathrm{mK}$ \\
$M_{\mathrm{AFM}}$ (max.\ mutual inductance for qubit pairs) & $1.647~\mathrm{pH}$ \\
Quantum critical point for 1D chains & $1.391~\mathrm{GHz}$ \\
$L_q$ (qubit inductance) & $372~\mathrm{pH}$ \\
$C_q$ (qubit capacitance) & $119~\mathrm{fF}$ \\
$I_c$ (qubit critical current) & $2.1~\mu\mathrm{A}$ \\
Average single-qubit temperature  & $0.198$ \\
Ferromagnetic-problem freezeout & $0.064$ \\
Single-qubit freezeout & $0.612$ \\
$\Phi^{\mathrm{i}}_{\mathrm{CCJJ}}$ (external flux on CCJJ at $s=0$) & $-0.621~\Phi_0$ \\
$\Phi^{\mathrm{f}}_{\mathrm{CCJJ}}$ (external flux on CCJJ at $s=1$) & $-0.717~\Phi_0$ \\
Readout time range & $17.0$ to $235.0~\mu\mathrm{s}$ \\
Programming time & $\sim 14100~\mu\mathrm{s}$ \\
QPU delay time per sample & $20.5~\mu\mathrm{s}$ \\
Readout error rate & $\le 0.001$ \\
\hline
\end{tabular}
\caption{Physical properties of the D-Wave Advantage 4.1 annealer.}
\label{tab:qpu-physical-properties}
\end{table*}

\end{document}